\documentstyle[aps,floats,prl,twocolumn]{revtex} 
\newcommand{\bm}[1]{\mbox{\boldmath $#1$\unboldmath}}
\begin{document} 
\title{Resonant Raman scattering by collective modes of the
one-dimensional electron gas.}
\author{Maura Sassetti$^{1}$ and Bernhard Kramer\\
I. Institut f\"ur Theoretische Physik, Universit\"at Hamburg\\
Jungiusstra\ss{}e 9, D-20355 Hamburg, Germany}

\author{
{\small (December 15, 1997)}
\vspace{3mm}
}
\author{%
\parbox{14cm}{\parindent4mm\baselineskip11pt%
{\small
We show that the low-energy peak in the polarized resonant Raman
spectra of quantum wires, which is commonly associated with ``single
particle excitations'', can be interpreted as signature of intra-band
collective spin excitations. A broad maximum in the resonant
depolarized spectra is predicted to exist above the frequency of
the spin density excitation, due to simultaneous but
independent propagation of spin- and charge-density modes.
}
\vspace{4mm}
}
}
\author{
\parbox{14cm} {\small PACS numbers: 71.45.-d, 73.20.Dx, 78.30.-j
}
}
\maketitle

Inelastic light scattering by etched AlGaAs/\-GaAs-quan\-tum wires at
Helium temperatures shows pronoun\-ced features that are commonly
inter\-preted in terms of inter- and intra-subband excitations of the
interacting electron gas. Depending on the relative polarization of the
incident and the scattered light, charge-density excitations (CDE,
parallel polarization, polarized spectra) and spin-density excitations
(SDE, crossed  polarization, depolarized spectra) have been identified.
Near resonance, when the frequency of the incident light is close to
the frequency of the band gap, additional structures have been detected
in both types of spectra which have been interpreted as being the
signature of ``single particle excitations'' (``SPE'')
\cite{gpwetal93,sgpetal94,setal94,sbketal96}. The intensities of these
peaks depend strongly on the frequency of the incident light. Their
physical origin is presently controversely discussed. In
\cite{gpwetal93,sgpetal94} where the electron density and the geometry
of the wires have been such that only two subbands were occupied, the
Fermi-liquid character of the 1D electron gas has been stressed. In
\cite{sbketal96}, with several subbands involved in the inelastic
scattering, the ``SPE'' have been interpreted as ``energy-density
fluctuations''. Recent results of very elaborate 
RPA calculations \cite{hs93} seem to
be consistent with the absence of ``SPE'' at low excitation energies in 1D.
By including two occupied subbands \cite{sh94}, the
observed ``SPE'' peaks at low energies have been interpreted as an
out-of-phase CDE in the two-band system in which the long-range part of
the Coulomb interaction is partially cancelled \cite{s93}. However, as has been
pointed out in \cite{gpwetal93}, the energies of these excitations,
as observed experimentally, seem to be much too small. 

In this paper, we point out that the peaks at low excitation energy in
the {\em polarized Raman spectra} which are strongly enhanced when the
photon energy approaches the energy gap, can be understood within the
theory of the collective excitations of the 1D electron gas with spin
in the Luttinger approximation. We show that they are signatures of the
collective SDE and appear at the same energy,
namely $\hbar v_{\rm \sigma }|q|$, as the SDE in the depolarized spectra
($v_{\rm \sigma }$, $q$ velocity and wave number of spin density
excitations, respectively). They appear due to higher order terms in
the Raman cross section related to the resonance condition. These terms
induce a relaxation of the selection rules valid for non-resonant Raman
scattering. Our results indicate for quantum wires a solution of the
puzzle of the ``SPE'' that has been posed already more than two decades
ago for inversion layers \cite{pbba71}. We quantify the suggestion of a
singlet spin mode being responsible for the ``SPE'' peaks which has
been mentioned to the best of our knowledge for the first time in
\cite{s93}.

We predict that, when $\hbar v_{\rm F}q/|E_{\rm G}-\hbar\omega _{\rm i}|\ll
1$ ($E_{\rm G}$ energy gap, $v_{\rm F}$ Fermi velocity),
the strength of the ``SPE'' peak varies with the energy $\hbar\omega _{\rm
i}$ of the incident light as
\begin{equation}
I_{{\rm SPE}}\propto |E_{\rm G} - \hbar\omega _{\rm i}|^{-4}\,,
\label{spe}
\end{equation}
in lowest order, and {\em increases} quadratically with temperature. As
a further consequence of the resonance condition, we predict that
structure associated with simultaneous propagation of spin- and charge
density excitations should appear in the {\em depolarized spectra}. In
contrast to the SDE in the polarized spectra, this
is not a sharp peak but merely a rather broad maximum in the scattered
intensity on the high energy side of the SDE-peak. 

Our results show that the Raman excitation spectra 
of quantum wires at low energies, in the region of {\em intraband 
transitions}, can be understood within the non-Fermi liquid 
framework of the 1D electron gas. 

Within the standard theory of Raman scattering \cite{b70} the differential 
cross section at $\omega=\omega_{\rm i}-\omega_{\rm f}$,
the difference between the frequency of the light in the initial 
and final state, is given by the average
\begin{equation}
\frac{{\rm d}^{2}\sigma }{{\rm d}\Omega {\rm d}\omega } \propto
	\left\langle \sum_{{\rm f}}| M_{{\rm fi}} |^{2}
	\delta (E_{{\rm f}}-E_{{\rm i}}-\hbar\omega )
				\right\rangle_{{\rm i}}
\end{equation}
where i and f denote initial and final states, respectively, and 
$\langle\cdots\rangle_{\rm i}$ is the thermal average over the initial
state. The transition matrix elements $M_{\rm fi}$ consist of terms
proportional to ${\bm A}^{2}$ and ${\bm \Pi}\cdot{\bm A}$ (${\bm A}$,
${\bm \Pi }$ vector potential and momentum operator, respectively). The
latter has to be treated in second order, and requires further
approximations, especially near resonance. The final result,
taking into account only one conduction and one valence band (with effective
masses $m_{\rm c}$ and $m_{\rm v}$, respectively),
can be written in terms of a generalized correlation function (${\bm
q}\equiv {\bm k}_{\rm i}-{\bm k}_{\rm f}$), 
\begin{equation}
\frac{{\rm d}^{2}\sigma }{{\rm d}\Omega {\rm d}\omega }=
	\left(\frac{e^{2}}{m c^{2}}\right)^{2}
	\frac{\omega _{\rm f}}{\omega _{\rm i}}
	\frac{n_{\omega }+1}{\pi }{\cal I}{\rm m}\chi ({\bm q},\omega )\,,
\end{equation}
with $\chi({\bm q},t)={\rm i}\Theta(t) \langle
\left[N^{\dagger}({\bm q},t),N({\bm q},0)\right]\rangle$, 
and the Bose distribution $n(\omega )$. The operator
\begin{equation}
N({\bm q})=\sum_{s,k}\frac{\gamma _{s}}{D({\bm k})}
	c_{s}^{\dagger}({\bm k}+{\bm q})c_{s}({\bm k})
\end{equation}
contains the Fermion operators $c_{s}^{\dagger}({\bm k})$,
$c_{s}({\bm k})$
with wave vector ${\bm k}$ and spin $s=\pm$. 
The coefficients are
\begin{equation}
\gamma _{s}=\gamma _{0}
	\left({\bm e}_{\rm i}\cdot {\bm e}_{\rm f}
	+{\rm i}s|{\bm e}_{\rm i}\times {\bm e}_{\rm f}|\right)\,.
\label{coefficients}
\end{equation}
Here, ${\bm e}_{{\rm i,f}}$ are the polarization vectors of the
incoming and outgoing electromagnetic fields, and
\begin{equation}
D({\bm k})=E_{\rm c}({\bm k}+{\bm q})-E_{\rm v}({\bm k}+{\bm q}
	-{\bm k}_{\rm i})-\hbar\omega _{\rm i}
\end{equation}
contains the energies of the valence and conduction bands $E_{\rm v}$
and $E_{\rm c}$ (in the effective mass approximation), respectively,
and the wave vector of the incoming light, ${\bm k}_{\rm i}$. The
prefactor $\gamma _{0}$ contains the matrix elements for the
transitions between the valence and the conduction band. It is assumed
to be constant in the following.

If $\hbar v_{\rm F}q\ll |E_{\rm G}-\hbar\omega _{\rm i}|$
and $\hbar \omega \ll |E_{\rm G}-\hbar\omega _{\rm i}|$, one can
neglect the ${\bm k}$-dependence of $D({\bm k})\approx E_{\rm
G}-\hbar\omega _{\rm i}$ (with $E_{\rm G}=E_{\rm G}^{0}+ \eta E_{\rm
F}$ $(\eta = 1+m_{\rm c}/m_{\rm v}))$. Here, $E_{\rm G}$ is the
distance between the conduction and the valence band at the
Fermi wave number $k_{\rm F}$ and $E_{\rm F}=\hbar^{2}k_{\rm
F}^{2}/2m_{\rm c}$. Then, the operator
\begin{equation}
N({\bm q})=\frac{\gamma _{0}}{E_{\rm G}-\hbar\omega _{\rm i}}
	\left[{\bm e}_{\rm i}\cdot {\bm e}_{\rm f}\,\rho ({\bm q})
	+{\rm i}|{\bm e}_{\rm i}\times {\bm e}_{\rm f}|
	\,\sigma ({\bm q})\right]
\label{operator1}
\end{equation}
is proportional to the charge density
$\rho ({\bm q})=\rho _{+}({\bm q})+\rho _{-}({\bm q})$
or to the spin density
$\sigma ({\bm q})=\rho _{+}({\bm q})-\rho _{-}({\bm q})$
depending on whether incoming and outgoing light are
polarized parallel or perpendicular, respectively. Thus, in lowest
order, one observes charge-density excitations in polarized, and 
spin-density excitations in the depolarized configuration. This is the
``classical'' selection rule of Raman spectra of quantum wires and dots.

Close to the resonance, when $\hbar \omega _{\rm i}\approx E_{\rm
G}+\hbar v_{\rm F}q$, the assumption of a constant energy denominator
is no longer valid. We expect that the above selection rule is relaxed.
This will now be shown for a quantum wire and expanding $D({\bm
k})^{-1}$ to first order in $\hbar v_{\rm F}q (E_{\rm G}-\hbar\omega
_{\rm i})^{-2}$. Specifically, we assume $E_{\rm c}=\varepsilon _{n}+
\hbar^{2}k^{2}/2m_{\rm c}$, with the subband energies $\varepsilon
_{n}$ ($n=0,1,2,3\ldots$) determined by the confinement in the $y$- and
the $z$-directions, and $k$ the wave number for the $x$-direction.
Assuming back scattering, $k_{\rm i}=q/2$, and considering only the
lowest subband,
\begin{equation}
D(k)=E_{\rm G}^{0} + \eta
\frac{\hbar^{2}k^{2}}{2m_{\rm c}} +
	\frac{\hbar^{2}kq}{m_{\rm c}}\xi
	-\hbar \omega _{\rm i} + {\rm O}(q^{2})\,,
\label{d(k)} 
\end{equation}
with $\xi = 1+m_{\rm c}/2m_{\rm v}$. Since $q\ll k_{\rm F}$ we can
linearize around $k=\pm k_{\rm F}$. Then, we can use the Luttinger model
\cite{l63,h81} in order to evaluate the correlation function $\langle
N^{\dagger}(q,t)N(q,0)\rangle$. In this model, it is useful to
introduce a decomposition of the energy spectrum into branches $b$ that
correspond to left- and right-moving excitations, $b=-$ and $b=+$,
respectively and $N(q)=\sum_{b}N^{(b)}(q)$. 

The expansion of the inverse of the energy denominator yields
contributions to $N(q)$ which are of the form of the ``energy density
fluctuations'' mentioned in \cite{w68},
\begin{eqnarray}
\Delta N^{(b)}(q)&=&
	-\frac{b\eta \hbar v_{\rm F}}
		{(E_{\rm G}-\hbar\omega _{\rm i})^{2}}
	\sum _{s,k}\gamma _{s}\cdot (k-bk_{\rm F})\nonumber\\
	&\times&c_{s}^{(b)\dagger}(k+q)c_{s}^{(b)}(k)\,.
\end{eqnarray}
These can be expressed by the above charge- and spin-density operators
by using the bosonization technique developped earlier for the
Luttinger model in \cite{h81,v95}.

After a straightforward calculation, one obtains
terms of the form (\ref{operator1}) but 
$\propto \hbar v_{\rm F}q/(E_{\rm G}-\hbar \omega _{\rm i})^{2}$
and additionally new contributions which are quadratic in the densities 
\begin{eqnarray}
&&\Delta N^{(b)}(q)=-\frac{\eta v_{\rm F}\gamma _{0}}
	{(E_{\rm G}-\hbar\omega _{\rm i})^{2}}
	\frac{\pi }{2L}\times\nonumber\\
&&\sum_{k}\left[2{\rm i}|{\bm e}_{\rm i}\times {\bm e}_{\rm f}|
	\rho ^{(b)}(k)\sigma ^{(b)}(q-k)
+({\bm e}_{\rm i}\cdot {\bm e}_{\rm f})\right.\nonumber\\
&&	\left. :\rho ^{(b)}(k)\rho ^{(b)}(q-k) +
	\sigma ^{(b)}(k)\sigma ^{(b)}(q-k):\right]\,,
	\label{deltan}
\end{eqnarray}
where the $:\cdots:$ stand for the normally ordered product 
of the operators.

Equation (\ref{deltan}) is the main result of this work. The evaluation
of the corresponding correlation function can be done exactly but is
considerably more complicated than for $N(q)\propto\sigma (q)$, for
instance. However, the form of $\Delta N(q)$ shows that in general the
spin density fluctuations will contribute to the cross section in the
polarized configuration besides the charge density fluctuations.
Correspondingly, signatures of the latter can be expected in the
depolarized spectrum. The ``classical'' selection rule which says that
charge-wave excitations appear only in the polarized configuration and
spin-wave excitations only in the depolarized spectrum, respectively,
is only valid in the lowest approximation, when the wave vector
dependence of $D(k)$ is neglected.
 
For the evaluation of the contribution of the spin excitations to the
correlation function of the polarized spectrum we need to calcu\-late 
correlators of the form
\begin{equation}
\langle [\sigma (k,t)\sigma (-q-k,t),\sigma
(k',0)\sigma (q-k',0)]\rangle\,,
\label{quartic}
\end{equation}
instead of the correlation function
$
\langle[\sigma (-q,t),\sigma (q,0)]\rangle
$
which determines the cross section in the depolarized configuration,
when the energy denominator is constant.

The calculation of the correlation functions
can be per\-for\-med by
using the Lut\-tin\-ger Hamil\-tonian $H = H_{\rho } + H_{\sigma }$, where
$H_{\rho }$ is the quad\-ratic form describing the charge density
excitations \cite{s93,s95,v95} and the spin part is
\begin{eqnarray}
H_{\sigma }&=&\frac{\pi\hbar v_{\rm F}}{L}\sum_{q>0}
	\left[\sigma ^{(+)}(q)\sigma ^{(+)}(-q)
	+ \sigma ^{(-)}(-q)\sigma ^{(-)}(q)\right]\nonumber\\
	  &&\qquad\qquad-\frac{g_{1}}{L}\sum_{q>0}
			\sigma ^{(+)}(q)\sigma ^{(-)}(-q)\,.
\label{spinhamiltonian}
\end{eqnarray}
The Hamiltonian can be diagonalized by a Bogolubov transformation. The
spectrum of the charge modes is
\begin{equation}
\omega _{\rho }(q)=v_{\rm F}|q|\left\{
\left(1+\frac{g_{1}}{h v_{\rm F}}\right)
\left[1-\frac{g_{1}}{h v_{\rm F}}+4\frac{V(q)}{h v_{\rm F}}\right]
\right\}^{1/2}\,,
\end{equation}
with $V(q)$ the Fourier transform of the interaction potential and
an interaction constant $g_{1}$ which describes a part of 
the exchange interaction.
For the 
spin density excitations
\begin{equation}
\omega _{\sigma }(q)=v_{\rm F}\sqrt{1-\frac{g_{1}^{2}}{h^{2}v_{\rm
F}^{2}}}\,|q|\equiv v_{\rm \sigma }|q|\,.
\end{equation}

The result of the calculation of the correlation functions valid {\em
near resonance} can be written in a closed form, but the remaining
integrals have to be computed numerically \cite{sk98}. For the present
purpose, it is sufficient to consider $\hbar v_{\rm F}q<<|E_{\rm G}
-\hbar \omega _{\rm i}|$ which gives for the spin contribution to the
polarized spectrum at the temperature $T$ a peak at the frequency
$v_{\sigma}q$,
\begin{eqnarray}
{\cal I}{\rm m}\chi (q,\omega )&=&
	\frac{Lq\,(\eta \hbar v_{\sigma}\gamma _{0})^{2}}
		{12\,(E_{\rm G}-\hbar \omega _{\rm i})^{4}}
	\left[\left(\frac{\pi k_{\rm B}T}
	{\hbar v_{\rm \sigma }}\right)^{2} +
	\frac{q^{2}}{2}\right]\nonumber\\
	&&\qquad\qquad\times\delta (\omega - v_{\rm \sigma }q)\,,
	\label{imaginarychi}
\end{eqnarray}
which corresponds to the same position in energy as that of
the SDE peak in the
depolarized spectrum {\em far from resonance},
\begin{equation}
{\cal I}{\rm m}\chi (q,\omega )=
	\frac{Lq\gamma _{0}^{2}}{(E_{\rm G}-\hbar \omega _{\rm i})^{2}}
	\delta (\omega - v_{\rm \sigma }q)\,.
\end{equation}

We note that although the peaks appear at the same energies, their
strengths depend differently on the photon energy. While the weight of
the SDE peak increases quadratically with increasing $|E_{\rm G} -
\hbar \omega _{\rm i}|^{-1}$, the peak in the polarized spectrum
increases with the 4th power. Also, the SDE-related peak in the
polarized spectrum far from resonance is independent of the
temperature, due to the linearization of the spectrum, while the peak
in the polarized spectrum increases quadratically with $T$.

In the depolarized configuration, we obtain also a relaxation of the
``classical'' selection rules near resonance. The cross section in next
higher order contains correlation functions of the form
\begin{equation}
\langle [\rho (k,t)\sigma (-q-k,t),
	\rho (k',0)\sigma (q-k',0)]\rangle\,.
\label{mixed}
\end{equation}
and no correlation functions with four charge density operators alone.
Due to the absence of spin-charge coupling in the Hamiltonian,
(\ref{mixed}) factorizes into products of the type $\langle\sigma
(-q-k,t)\sigma (q-k',0)\rangle \langle \rho (k,t)\rho (k',0)\rangle$
indicating independent motion of the spin and charge modes. Due to the
presence of terms like (\ref{mixed}), we do not expect structure in the
cross section that is solely determined by the charge density
excitations. Indeed, we find that the simultaneous propagations of the
two types of excitations leads to a broad continuum in the depolarized
spectrum above the frequency $\omega _{\sigma }(q)$.

Also in the contributions that are still higher order in $\hbar v_{\rm
F}q (E_{\rm G} - \hbar \omega _{\rm i})^{-2}$, we do not find correlators
that contain only charge density operators since the depolarized part
of the cross section originates in the spin-orbit coupling (\cite{b70},
cf. (\ref{coefficients})) and the corresponding excitation processes
are accompanied by spin-flip processes. This implies that all of the
terms contributing towards the depolarized cross section must contain
at least one pair of spin density operators, and structure related to
the charge density excitations alone is absent.
 
Comparing with experiment, we first note that all works agree 
in the linear
dependence of the excitation energy on the wave number of the peak
associated with the ``SPE'' in the polarized spectrum. In
\cite{sbketal96}, the velocity of the ``SPE'' has been found to be
approximately the same as the velocity of the SDE determined from the
depolarized spectra, and approximately equal to the Fermi velocity in
the lowest occupied subband. Our results are consistent with this, if
we assume that $g_{1}/h v_{\rm F}\ll 1$. If the Fermi velocity was
determined independently, the spin interaction constant $g_{1}$ could
in principle be determined. However, it is expected that $g_{1}$ is in
any case very small \cite{s93} so that $v_{\rm \sigma }= v_{\rm F}$ to
a very good approximation. In \cite{gpwetal93}, data have been
presented (Figs.~1 and 2)  which seem to indicate that ``SPE'' and SDE
peaks are slightly different in energy, the velocity of the former
being approximately $v_{\rm F}$ while that of the latter has been
identified to be slightly smaller. However, when taking the error bars
into account it is not possible to distinguish between the positions of
the peaks. Thus, these results can also be considered to be consistent
with our present model.

On the other hand, we cannot exclude that there are
many-particle corrections to the spin excitations beyond our model. 
For instance, if the spin Hamiltonian (\ref{spinhamiltonian}) contained
additional quartic terms, the poles of the above quartic correlators
(\ref{quartic}) would be different in energy from those of the
correlators quadratic in the spin density. Also, corrections due to
higher subbands could lead to different velocities of low-energy
excitations. However, the recent experiments on samples with several
subbands occupied \cite{sbketal96}, show that this is very improbable.

To the best of our knowledge, there are up to now no systematic
measurements of the dependence of the heights and the widths of the
``SPE'' peaks as functions of the photon energy and the temperature.
Such measurements could provide further support for our interpretation.
One should have in mind that the {\em precise} value of the gap energy
$E_{\rm G}$ is not known. Measurement of the dependence on the photon
energy of the incoming light would provide the possibility of
determining $E_{\rm G}$. Closer to resonance, approximation
(\ref{spe}) is insufficient. Here, one has to evaluate
numerically the $k$-integral in the correlator without expanding
$D(k)^{-1}$ \cite{sk98}.

Concerning the additional structure in the depolarized
spectrum predicted above, we could find only very weak experimental 
evidence in Fig.~2 of \cite{gpwetal93}. 
These authors interprete a
slight asymmetry in the peak associated with SDE as a signature of the
``SPE''.  Our findings offer a different
interpretation: the asymmetry could be due to the continuum
contribution to the depolarized spectrum which originates in the
motions of simultaneouly excited spin- and charge-density waves.
However, further experiments using wires with only one subband occupied are
necessary, in order to confirm or to disprove this interpretation. 
 
Within the present model, we cannot comment on the experimental {\em
interband} results in quantum wires with higher subbands involved,
where ``SPE'' and SDE have clearly different excitation energies.  Due
to the comparatevely high excitation energies, the pecularities of the
Luttinger model are absent in this region. Especially, one can expect
the Fermi-liquid character of the electron gas to be restored. However,
we also expect for these excitations corrections towards the Raman
cross sections in both configurations due to wave vector dependent
terms in $D(k)$.  We suspect that these (i) do not obey the
``classical'' selection rules and (ii) will in general produce
structures at different energies than those of SDE and CDE. 

In summary, we have presented results for the intra-band Raman spectra
of a quantum wire with only one subband occupied. They are consistent
with all of the experimental findings presently available at low
excitation energies. We have shown that the low-energy ``SPE'' in the
{\em polarized spectrum} near resonance can be interpreted as signature
of the spin-density excitations of the 1D electron gas. When accepting
this, the presently available data of resonant Raman scattering can be
taken as indicating the charge-spin separation predicted by
the Luttinger model and for the non-Fermi liquid
character of the 1D electron gas at low excitation energies.

The measurement of the above predicted dependence of the peak
intensities on the photon energy and on the temperature, namely
$\propto |E_{\rm G}-\hbar\omega _{\rm i}|^{-4}$ and $T^{2}$,
respectively, could further confirm our interpretation. In addition, we
predict near resonance a continuum in the cross section which extends
above the frequency of the spin excitations in the {\em depolarized
spectrum}. It is related to simultaneous but independent propagation of
spin and charge modes. 

We gratefully acknowledge useful discussions with Detlef Heitmann, and 
Christian Sch\"uller. The work has been performed within the TMR
networks FMRX-CT96-0042 and ERB4061-PL970066 of the European Union;
financial support has been obtained from the Deut\-sche
For\-schungs\-ge\-mein\-schaft via SFB 508
``Quan\-ten\-ma\-te\-ria\-lien'', the Gra\-duier\-ten\-kolleg
``Nano\-struk\-tu\-rier\-te Festk\"orper'', Project Kr 627 and from
Istituto Nazionale di Fisica della Materia within PRA97(QTMD).

$^{1}$ on leave of absence from  Istituto di Fisica di Ingegneria,
INFM, Universit\`a di Genova, Via Dodecaneso 33, I--16146 Genova.

\end{document}